\documentclass[preprint]{aastex}

\usepackage{float}
\usepackage{amsmath}
\usepackage{graphics,graphicx}
\usepackage{natbib}

\shorttitle{Fe K$\alpha$ Profiles from Simulations}
\shortauthors{Kinch et al.}

\begin{document}

\title{Fe K$\alpha$ Profiles from Simulations of Accreting Black Holes}

\author{Brooks E. Kinch}
\affil{Department of Physics and Astronomy, Johns Hopkins University, Baltimore, MD 21218, USA}
\email{kinch@jhu.edu}

\author{Jeremy D. Schnittman}
\affil{NASA Goddard Space Flight Center, Greenbelt, MD 20771, USA}
\email{jeremy.d.schnittman@nasa.gov}

\author{Timothy R. Kallman}
\affil{NASA Goddard Space Flight Center, Greenbelt, MD 20771, USA}
\email{timothy.r.kallman@nasa.gov}

\and

\author{Julian H. Krolik}
\affil{Department of Physics and Astronomy, Johns Hopkins University, Baltimore, MD 21218, USA}
\email{jhk@jhu.edu}

\begin{abstract}

We present first results from a new technique for the prediction of Fe K$\alpha$ profiles directly from general relativistic magnetohydrodynamic (GRMHD) simulations. Data from a GRMHD simulation are processed by a Monte Carlo global radiation transport code, which determines the X-ray flux irradiating the disk surface and the coronal electron temperature self-consistently. With that irradiating flux and the disk's density structure drawn from the simulation, we determine the reprocessed Fe K$\alpha$ emission from photoionization equilibrium and solution of the radiation transfer equation.  We produce maps of the surface brightness of Fe K$\alpha$ emission over the disk surface, which---for our example of a $10 M_\odot$, Schwarzschild black hole accreting at $1\%$ the Eddington value---rises steeply one gravitational radius outside the radius of the innermost stable circular orbit and then falls $\propto r^{-2}$ at larger radii.  We explain these features of the Fe K$\alpha$ radial surface brightness profile as consequences of the disk's ionization structure and an extended coronal geometry, respectively.  We also present the corresponding Fe K$\alpha$ line profiles as would be seen by distant observers at several inclinations. Both the shapes of the line profiles and the equivalent widths of our predicted K$\alpha$ lines are qualitatively similar to those typically observed from accreting black holes. Most importantly, this work represents a direct link between theory and observation: in a fully self-consistent way, we produce observable results---iron fluorescence line profiles---from the theory of black hole accretion with almost no phenomenological assumptions.

\end{abstract}

\keywords{accretion, accretion disks -- black hole physics -- line: formation -- line: profiles -- radiative transfer -- X-rays: binaries}

\section{Introduction}

The immediate environment of black holes offers a unique laboratory for astrophysics.   Accretion onto black holes is among the most efficient of astrophysical processes---it is the mechanism behind their prodigious luminosity.   In addition, black holes provide an opportunity to explore the strong-field regime ($GM/Rc^2 \sim 1$, where $M$ and $R$ are the mass and characteristic scale of the gravitational source, respectively) of general relativity.

The Fe K$\alpha$ line can give us valuable information about this regime.   Due to the relatively high abundance of iron in the Universe and its efficient production of photons by fluorescence and recombination, a strong emission line can be created whenever a strong hard X-ray source is present.   In addition, a relativistically broad profile is an immediate signal of an origin deep in the black hole's potential.   Indeed, relativistically broadened Fe K$\alpha$ lines have been observed in both AGN \citep{tan95a, nan07a, bre09a} and galactic black holes \citep{mil04a, rei08a, rei09a}, as well as galactic neutron stars \citep{cac10a}.

Detailed observations of this line therefore offer a direct channel to dynamics in the strong-field regime of general relativistic gravity.  In principle, its flux, energy profile, and variability properties could constrain many aspects of accretion dynamics, disk coronae, and possibly general relativity itself [e.g., Fe K$\alpha$ reverberation-mapping can lead to scale estimates: \citet{kara15} and references therein].    The shape of the K$\alpha$ profile has already been used as a tool with which to measure black hole spin  [see \citet{rey13a} and \citet{mil15a} for reviews], and several dozen individual measurements have been reported.    However, interpretation of such data to date has in general relied upon ``phenomenological'' models.   For example, spin parameter inferences often rely on assumed functional forms of the Fe K$\alpha$ surface brightness, typically decreasing power-laws in the radial coordinate with some inner radius cutoff [see \citet{rey03a} for a review]. At best, these assumptions introduce additional parameters, but it is also possible that the chosen functional forms do not resemble the actual surface brightness profile. {  Many models (which we discuss below) \emph{do}, in fact, perform radiation transfer and photoionization calculations; these form the class of disk reprocessing codes, like that which we present in this paper. These more sophisticated models nonetheless must still make arbitrary assumptions about important physical quantities such as the spatial-dependence of the X-ray flux striking the disk and the internal density structure of the matter within the disk.}

These difficulties are well-illustrated by the common assumption that the inner cutoff of K$\alpha$ emission falls precisely at the radius of the innermost stable circular orbit (ISCO).  As the radius of the ISCO is a one-to-one function of the central black hole spin, the radius of disk truncation enters into parametric models as a proxy for the spin. But does the disk---and, presumably, the Fe K$\alpha$ emission---\emph{in fact} truncate right at the radius of the ISCO \citep{rey97a, kro02a}? Matter which has passed through the ISCO must still travel to the event horizon, so the density is necessarily nonzero there---and perhaps the Fe K$\alpha$ surface brightness is as well. Moreover, as the inward radial acceleration begins outside the ISCO \citep{kro05a}, the fall in surface density also starts there; the cutoff in Fe K$\alpha$ surface brightness might then likewise be found outside the ISCO.   Thus, to make Fe K$\alpha$ a {\it quantitative} diagnostic, it is essential to understand better the physical processes controlling its emission.

To this end, we have built numerical machinery to predict Fe K$\alpha$ emission in a way that is as close to first principles and as free from arbitrary assumptions and parameters as possible.   In this paper we show how it is possible to go directly from the data generated by a general relativistic MHD accretion simulation to observed K$\alpha$ profiles using {\it only} well-known physics.   As a proof of principle for the methods we describe below, we present the results for a single example: a $10 M_{\odot}$ Schwarzschild black hole accreting at $1 \%$ of the Eddington rate.

\section{From First Principles to Fe K$\alpha$ Line Profiles}

The simple model of accretion onto a black hole consists of an optically thick, geometrically thin disk, above and below which is a diffuse, hot corona \citep{lia79a, haa91a}. Thermal photons emitted from the disk surface are boosted to high energies via inverse Compton scattering with mildly relativistic electrons in the corona. Some of these upscattered photons re-impinge on the disk surface, and those with sufficient energy---above the Fe K-edge at approximately 7.0 keV---eject inner shell electrons from iron atoms. As higher energy electrons fall to the now vacant lower energy levels, fluorescent photons of energy ranging from 6.4 keV to 7.0 keV, depending on iron ionization state, are emitted, forming the characteristic and prominent Fe K$\alpha$ emission line frequently observed in the X-ray spectra of such objects. The shape and strength of the iron line is a function of how the Fe K$\alpha$ emission varies over the disk surface, which in turn depends on the disk's thickness and vertical density profile, its temperature structure, the shape and intensity of the irradiating flux, and the iron abundance.

\subsection{The Disk Structure and the Irradiating Flux}

Our calculation begins with the general relativistic three-dimensional magnetohydrodynamic code {\sc harm3d} \citep{nob09a}. {\sc harm3d} is an intrinsically conservative GRMHD code which yields dynamic, three-dimensional information about the fluid density, four-velocity, magnetic pressure, gas pressure, and cooling at every point throughout the computational domain. To ensure the production of a geometrically thin disk---in the sense that its aspect ratio, $H_{\text{dens}}/r$ (where $H_{\text{dens}}$ is the density-weighted scale height), remains small---{\sc harm3d} solves a modified stress-energy conservation equation: in gravitationally-bound gas above a target temperature $T_*$, the excess heat is radiated away on an orbital timescale. The target temperature $T_*$ is chosen so as to achieve a target aspect ratio. Thus, at this point, we require only a target aspect ratio and the dimensionless spin parameter. For what follows, we consider one representative snapshot of a high-resolution simulation with $H_{\text{dens}} = 0.06$, $a/M = 0$ \citep{nob11a}, and at time $t = 12500M$, when the disk is nearly in a statistically steady-state.

\begin{figure}[H]
\begin{center}
\includegraphics[width=0.8\textwidth]{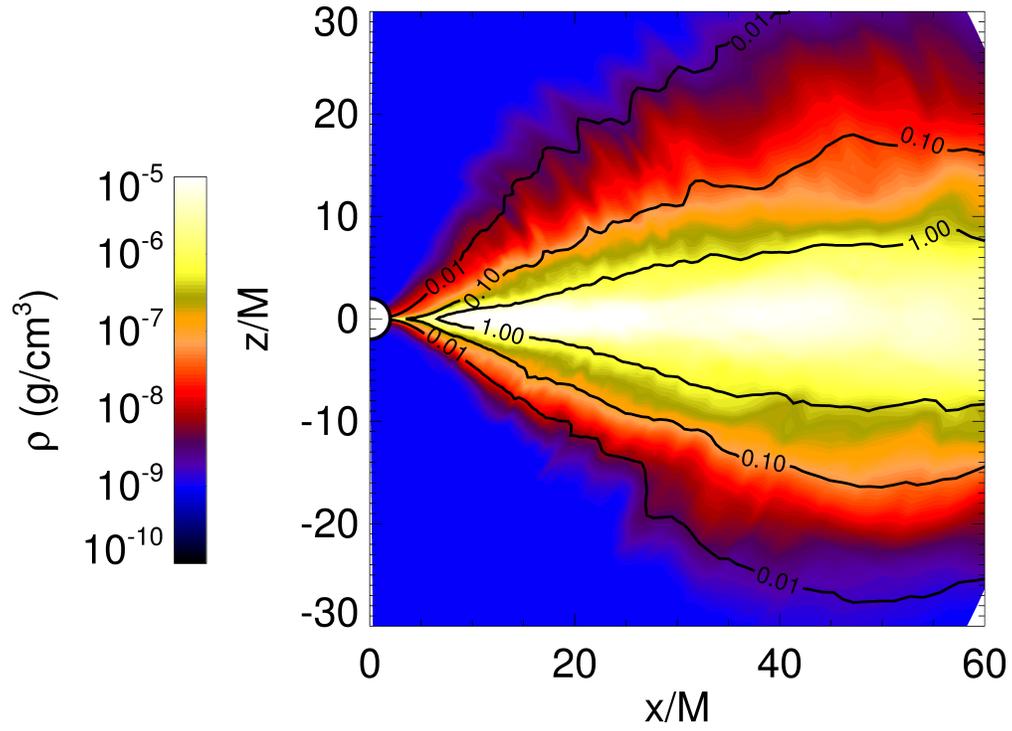}
\caption{The $\phi-$averaged gas density drawn from {\sc harm3d} at $t = 12500M$, scaled to $10 M_\odot$ and $\dot{m} = 0.01$ as described in the text, with several surfaces of constant optical depth overlaid.}
\label{rho_phi_avg}
\end{center}
\end{figure}

\begin{figure}[H]
\begin{center}
\includegraphics[width=0.8\textwidth]{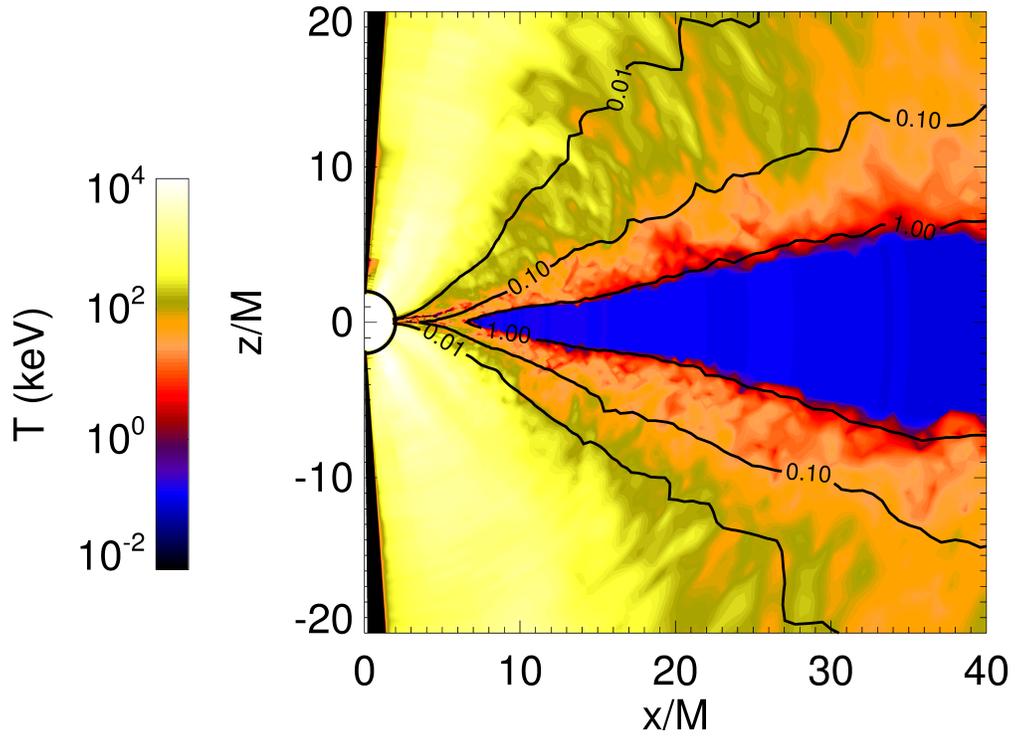}
\caption{The $\phi-$averaged temperature computed by {\sc pandurata}, with several surfaces of constant optical depth overlaid, computed as described in the text.}
\label{temp_phi_avg}
\end{center}
\end{figure}

\begin{figure}[H]
\begin{center}
\includegraphics[width=0.8\textwidth]{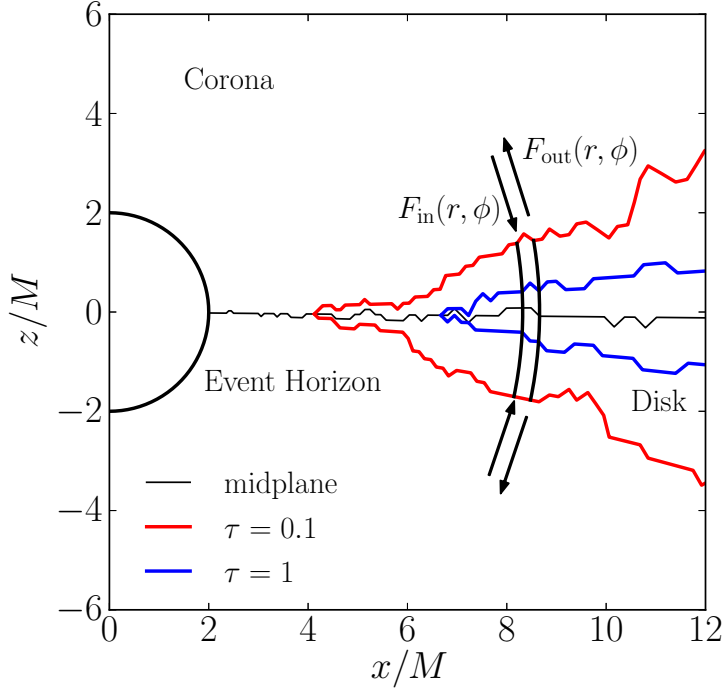}
\caption{A schematic, cross-sectional view of the black hole environment. For a Schwarzschild black hole, the event horizon is at $r = 2M$, and the ISCO (not labeled here) is at $6M$ in the equatorial plane. The red and blue lines indicate the $\tau = 0.1$ and $\tau = 1$ surfaces of constant optical depth; the thin black line indicates the midplane. The region outside the event horizon and the disk is the corona. The pair of solid black lines connecting the upper and lower $\tau = 0.1$ surfaces represent one of the finite columns in which we perform our transfer and photoionization solution. The arrow labeled $F_{\text{in}}(r, \phi)$ (and its counterpart for the lower surface) indicates the (energy-dependent) flux incident upon the disk at that $(r, \phi)$, computed by {\sc pandurata} as described in the text. The arrow labeled $F_{\text{out}}(r, \phi)$ indicates the reprocessed emission (e.g., Fe K$\alpha$ photons) computed by the code described in this paper.}
\label{schematic}
\end{center}
\end{figure}

The next step is to scale the simulation results from dimensionless code units to physical (i.e., cgs) units, a procedure summarized in \citet{sch13a}---this requires the further specification of a central black hole mass $M$, which sets the natural length and time scales (since we define $G = c = 1$, both time and space are measured in units of the black hole mass $M$; for a $10 M_{\odot}$ black hole, $1M = 1.5 \times 10^6 \ \text{cm}$), and the accretion rate $\dot{M}$, which sets the scale for the gas density, cooling rate, and magnetic pressure. We then construct the photosphere surfaces, which separate the disk from the upper and lower coronae, by integrating the optical depth $d\tau = \kappa \rho(r, \theta, \phi) r d\theta$, where $\kappa$ is the Thomson scattering opacity, at constant $(r, \phi)$ from the poles at $\theta = 0, \pi$ toward the $x-y$ plane. The upper and lower photospheres are those surfaces, $\Theta_{\text{top}} (r, \phi)$ and $\Theta_{\text{bot}} (r, \phi)$, respectively, for which the integrated optical depth reaches unity; the midplane, $\Theta_{\text{mid}} (r, \phi)$, is that surface for which the integrated optical depths from both poles are equal---it is typically close to $\pi/2$, but varies in space and time. Surfaces of constant optical depth can be constructed for other values of $\tau$---of particular relevance to our transfer solution is the $\tau = 0.1$ surface. {  We consider the region between the $\tau = 0.1$ and $\tau = 1$ surfaces to be a ``boundary layer'' between the disk and corona (see below).} A schematic illustrating how these surfaces relate to one another and to the geometry of a Schwarzschild black hole is presented in Figure \ref{schematic}. {  Figures \ref{rho_phi_avg} and \ref{temp_phi_avg} show $\phi-$averaged density and temperature maps, respectively, for the snapshot of the simulation we consider here, with several surfaces of constant optical depth overlaid. The density snapshot illustrates the sharp vertical density gradient associated with the disk's small scale height.  It also shows the location of the photospheric surface; even after azimuthal-averaging, there is still significant irregularity in this surface, the result of the large amplitude turbulence within the disk.  The temperature snapshot demonstrates how poor an approximation it is to think of the corona as a single zone: the temperature ranges from $\sim 1$~Mev in the nearly hollow cone along the polar axis to only a few keV just outside the disk photosphere.}

\begin{figure}[H]
\begin{center}
\includegraphics[width=0.8\textwidth]{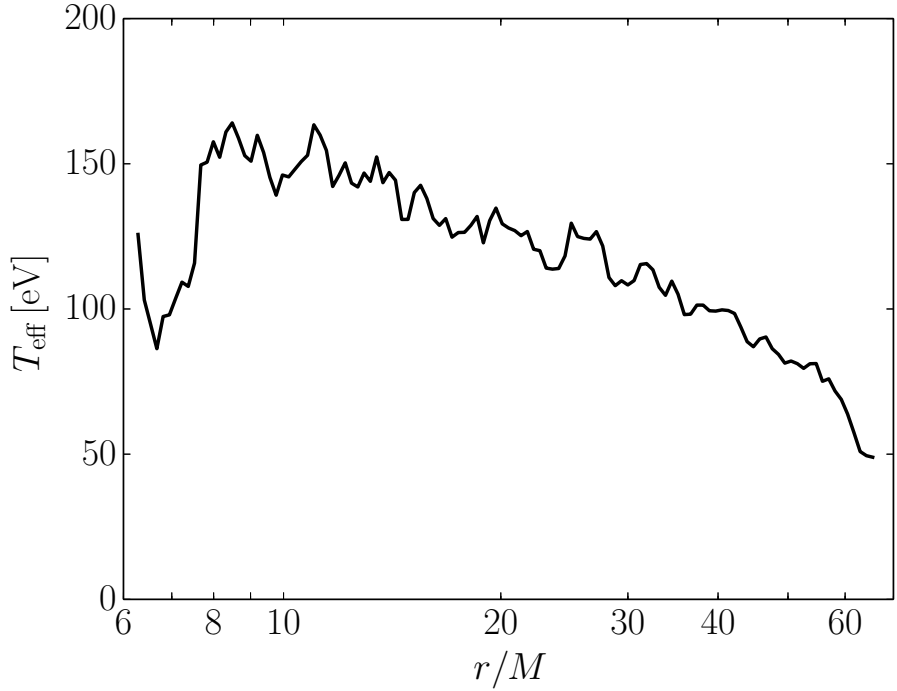}
\caption{The $\phi-$averaged effective temperature at the disk photosphere.}
\label{t_eff_phi_avg}
\end{center}
\end{figure}

In order to determine the shape and intensity of the hard X-ray flux incident upon each point of the disk surface, we employ {\sc pandurata}, a Monte Carlo relativistic radiation transport code \citep{sch13b}. {\sc pandurata} launches thermal seed photons from the photosphere and follows their trajectories through the curved spacetime around the black hole. These seed photons have a hardened blackbody energy distribution at a temperature consistent with the local cooling within the disk body as determined by {\sc harm3d}. That is, $T_{\text{eff}}$ at the photospheres $\Theta_{\text{top}}$ and $\Theta_{\text{bot}}$ at a given $(r,\phi)$ is set by assuming that the energy dissipated between them is radiated thermally at each photosphere equally \citep{sch13a}---i.e., $\int_{\Theta_{\text{bot}}}^{\Theta_{\text{top}}} d\theta \sqrt{g_{\theta\theta}} \mathcal{L} = 2 \sigma T_{\text{eff}}^2$. {  The $\phi-$averaged effective temperature at the photosphere, as a function of radius, is shown in Figure \ref{t_eff_phi_avg}. For most of the disk, the effective temperature declines slowly with radius, $\propto r^{-1/3}$, steepening to $\propto r^{-3/4}$ at large radii.} Some of these thermal seed photons escape to infinity (making up part of the observed spectrum), while some are lost to the black hole; others re-impinge on the disk surface, having been upscattered by relativistic electrons in the corona.

\begin{figure}[H]
\begin{center}
\includegraphics[width=0.8\textwidth]{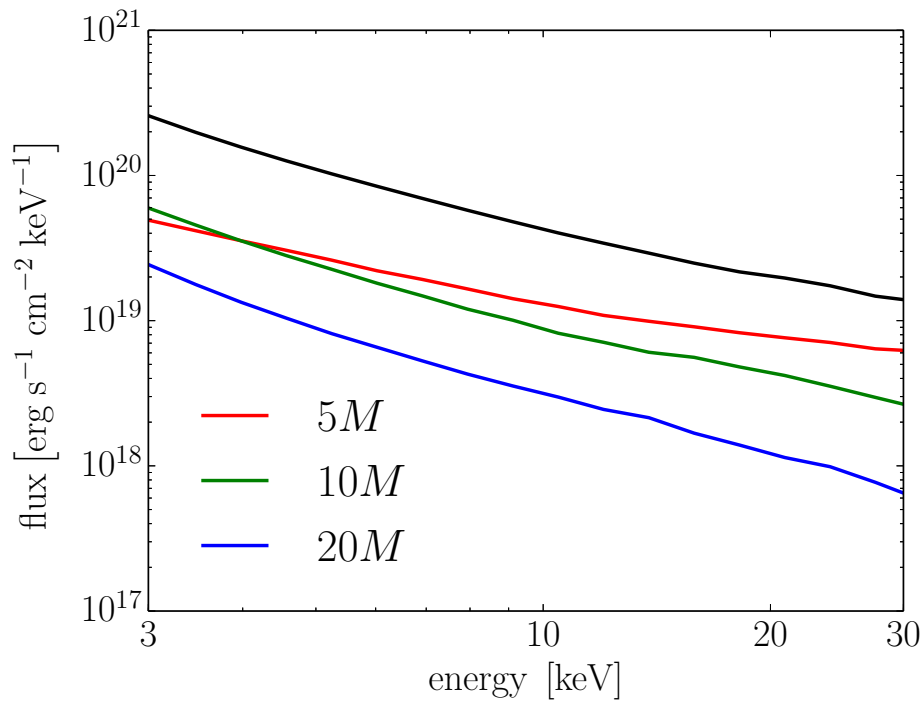}
\caption{The $\phi-$averaged X-ray flux incident upon the disk as a function of energy at three sample radii, compared to the flux in the same range as seen by a distant observer at $i = 45^\circ$. The scaling for the black curve is arbitrary, and is placed above the other three to facilitate comparison.}
\label{f_inc_vs_obs}
\end{center}
\end{figure}

In {\sc pandurata}, as photons traverse the corona, they have a chance to Compton scatter according to the opacity along their route. When a photon does scatter, the electron's velocity in the local fluid frame is chosen according to a relativistic thermal velocity distribution corresponding to the electron temperature at that point.  The photon's new direction is chosen from the Compton scattering partial cross section, and its new energy is then computed using standard relativistic dynamics. The new photon 4-momentum is then transformed back into the global coordinate frame. Eventually, a sufficient number of such events have occurred throughout the corona to permit evaluation of the inverse Compton (IC) power [the dominant emission mechanism in the corona \citep{sch13a}] at each coronal cell by directly comparing the incoming and outgoing photon energies. This value for the IC power at each coronal cell is then compared to the cooling rate found there by {\sc harm3d}, and the electron temperature (see Figure \ref{temp_phi_avg}) is adjusted so that the former better matches the latter---and so on, until ultimately a self-consistent picture of the coronal radiation field emerges, including the hard X-ray component incident upon the disk surface. {  This flux is shown, at several radii, in Figure \ref{f_inc_vs_obs}, compared to the spectrum as seen by a distant observer over the same range. The incident flux is well described by a power-law in energy with index ranging from $-0.8$ to $-1.8$, steepening at larger radii. The flux as seen by the distant observer is also a power-law in energy, with index $-1.4$, nearly independent of inclination. Thus, at small radii, the incident spectrum is somewhat harder than observed, while at larger radii ($r \gtrsim 15M$) it is somewhat softer than observed [cf. \citet{fur15a}]. See \citet{sch13a} for more details on the observed continuum spectra.}

Thus, having specified only the mass $M$, spin $a$, accretion rate $\dot{M}$, a list of elemental abundances, and a fiducial aspect ratio to ensure a geometrically thin disk, we construct a disk with a known density structure and a known flux irradiating its surface.

\subsection{The Transfer Solution}

To treat the detailed transfer of radiation and line photon production within the disk, we developed a new disk radiation reprocessing code---which we call {\sc ptransx}---coupled to the subroutines of the photoionization code {\sc xstar} \citep{kal01a} that calculate the local temperature, ionization, emissivity, and opacity. {\sc xstar}'s built-in transfer scheme is insufficient for our purposes for two reasons: it does not include scattering, which is an important effect in determining the effective Fe K$\alpha$ yield (see below), and it does not allow for a slab illuminated on \emph{both} sides, which---at small radii where the disk's total optical depth falls below order unity---has important consequences for the transfer solution. Instead, we employ Feautrier's method [see \citet{mih78a, mih85a} for a description of the method and numerical algorithm used here] applied to finite plane-parallel slabs, each of which is the region between the upper and lower $\tau = 0.1$ surfaces (see Figure \ref{schematic}) at fixed $(r, \phi)$. That is, we take for the density structure in one slab the (curved) column of cells from one snapshot of the GRMHD simulation with the same $(r, \phi)$ but with $\theta$ increasing from its value at the upper $\tau = 0.1$ surface to its value at the lower $\tau = 0.1$ surface (as in Figure \ref{schematic}).

{  The inclusion of a boundary layer just outside the disk photosphere is necessary in order to capture Fe K$\alpha$ emission in that region (see Figure \ref{yield}). The transition from corona to disk body is not sharp: in the fully-ionized, optically thin corona, Compton scattering is the dominant physical process; in the less-ionized, optically thick disk body, atomic absorption and emission become important as well. In the boundary layer, we use {\sc pandurata} to make a proper determination of the temperature, but {\sc ptransx} to determine the region's contribution to the Fe K$\alpha$ emission. The choice of $\tau = 0.1$ ensures both that all K$\alpha$ emitting Fe atoms are included in the transfer solution \emph{and} that the region which is \emph{not} treated by {\sc ptransx}, but by {\sc pandurata} alone, has no significant absorption opacity. We verify this \emph{post hoc}: over a wide range in energy, we find that the ratio of the absorption to scattering optical depths between the $\tau = 0.1$ and $\tau = 0.2$ surfaces is less than unity (typically $\lesssim 0.2$). Because the region between the poles and the $\tau = 0.1$ surface is even hotter, the ratio of absorption opacity to scattering must be even smaller there. So long as the volume in which the transfer solution is carried out contains the vast majority of fluorescing Fe atoms, the choice of which \emph{specific} surface of constant optical depth we choose is not important---choosing $\tau = 0.9$, for example, would yield similar results, but at the cost of extending {\sc ptransx}'s calculation into the region better suited to {\sc pandurata}.}

The relevant boundary conditions---the specific intensity incident upon the disk surface at its upper and lower boundaries---are taken from {\sc pandurata}, assuming isotropy of the incident radiation in the half-spaces immediately above and below the $\tau = 0.1$ surfaces. The transfer solutions in each $(r, \phi)$ slab are independent, and we justify this approximation by noting that the ionization parameter near the disk surface varies relatively slowly in space except for a region between $12M-14M$ in radius; this annulus, however, accounts for only a few percent of the total Fe K$\alpha$ emission. In addition, we set all elemental abundances, including iron, to their solar values---that is, we set $A_\text{Fe} = 7.50$ \citep{gre96a}.

At each point within the disk body, we define a floor temperature as $T_{\text{floor}} = T_{\text{eff}} \tau^{1/4}$ \citep{mih78a}. Above and below the photospheres, we scale the floor temperature as $T_{\text{floor}} = T_{1} \tau^{-3/4}$ \citep{sch13a}, where $T_{1}$ is the temperature found by {\sc pandurata} just outside the photosphere. If the temperature corresponding to photoionization equilibrium (as computed by {\sc xstar}) exceeds the floor temperature at any point, we use the photoionization equilibrium temperature; otherwise, we use the floor temperature. This scheme is valid, of course, only if a disk photosphere exists for which we might specify a value for $T_{\text{eff}}$. At small radii for which no photosphere exists---i.e., where the disk is optically thin, so defining an effective temperature by assuming radiation-gas thermal equilibrium is not physically reasonable---we simply adopt the temperature corresponding to photoionization equilibrium, with no floor. While the floor temperature is invoked within the disk body for nearly every slab at $r \gtrsim 10M$, removing the floor temperature entirely affects the Fe K$\alpha$ yield by about only $10\%$ (at least in this particular case).

We discretize each slab into cells such that the Thomson scattering optical depth, defined at the boundary of each cell, increases logarithmically from its value at the upper surface to its maximum value at the midplane, and likewise for the lower surface to the midplane. Further, we discretize $\mu$, the cosine of the angle with respect to the local plane normal (i.e., the $\hat{\theta}$ unit vector), into evenly-spaced bins between $-1$ and $1$, and the energy into logarithmically spaced bins between 1 eV and 500 keV; additional, evenly-spaced energy bins are added to the $6.3 - 7.0$ keV region in order to resolve Fe K$\alpha$ emission. For the analysis we consider below: 128 optical depth cells are spaced such that $\Delta \tau / \tau \simeq 0.01 - 0.09$, nearer to the lower value except for $(r, \phi)$ slabs at least several $M$ exterior to the ISCO; we employ 16 angle groups; and 1000 energy bins are spaced such that $\Delta E / E \simeq 0.013$, with 700 additional bins in the $6.3 - 7.0$ keV range. Numerical experimentation shows that the relevant results of our calculations are not appreciably affected by increasing the resolution beyond the values quoted above (for our purposes, in fact, far fewer angle groups are needed). Finally, we assume isotropic and coherent scattering.

We determine a self-consistent internal radiation field in the following way. Initially, we assume the gas to be completely ionized, so that there is zero true absorption. The scattering opacity at each point is known, given the density and corresponding electron number density. We then solve the equation of radiative transfer for the specific intensity as a function of angle, energy, and position. This procedure yields a value for the energy-dependent mean intensity at each point, which, along with the elemental abundances, density, and temperature floor, is supplied to the {\sc xstar} subroutines. These then return local values for the energy-dependent continuum absorption and continuum emissivity (including thermal bremsstrahlung, radiative recombination, and two-photon decays of metastable levels), line emissivity (both continuum and line emission are treated as isotropic), the free electron fraction, and photoionization equilibrium temperature (so long as it exceeds the floor, if supplied). From these values, we update the source function and opacities---and then re-solve the equation of radiative transfer, the mean intensity from which is again input for {\sc xstar}. We iterate until our quantity of interest---the Fe K$\alpha$ emission---has converged to within $1\%$; this typically requires $\sim 10$ iterations. In this fashion we arrive at a self-consistent radiation field, ionization balance, and temperature throughout the slab.

The computation of Fe K$\alpha$ emission is more detailed. The local emissivity due to a very large number of bound-bound transitions is computed by {\sc xstar} as part of the iterative solution described above. In the normal course of the solution, the line emissivities are binned with the continuum emissivities when computing the source function at each point---the outgoing flux determined this way includes both continuum and line contributions. After the convergence criterion is met, the Feautrier method is employed once again, but with the source function including \emph{only} continuum emission. In the $6.3-7.0$ keV range, line emissivity greatly exceeds continuum emissivity virtually everywhere in the disk; the outgoing continuum in this range is due to the reflected (or transmitted, at small radii) incident flux. All other quantities---e.g., the continuum scattering and absorption opacities---are those determined self-consistently with the \emph{full} radiation field. This final step yields the outgoing continuum flux, which is then subtracted from the outgoing total flux to arrive at the line emission part only, with no continuum-fitting necessary. We ignore the resonant absorption of Fe K$\alpha$ lines on the basis that their escape probabilities---computed by {\sc xstar} [see the {\sc xstar} documentation \citep{kal01a} and references therein]---are all very near to unity for the parameters considered here. The escape probability calculation requires knowledge of the local turbulent velocity of the gas, which we take, fiducially, as 5\% the Keplerian orbital velocity (a figure consistent with the disk's aspect ratio). However, the statement that the escape probabilities for line photons in the vicinity of the Fe K$\alpha$ line are nearly unity remains true over a wide range of turbulent velocities relative to the orbital velocity. Since the opacity for resonant absorption is small and such events are rare, we also safely ignore Auger destruction \citep{ros96a, kal04a}.

The end result of this calculation is that---starting from a small number of assumptions and working from first principles---we arrive at a self-consistent model of the line emission over the disk surface. In what follows, we restrict our attention to the Fe K$\alpha$ emission. As the energy of the K$\alpha$ emission varies with Fe ionization state, from 6.4 keV in neutral Fe to 7.0 keV in H-like Fe \citep{kro87a}, and because there are small but not completely negligible contributions from other heavy elements in this range, what we refer to as \emph{the} Fe K$\alpha$ flux is the sum of all emitted line photon fluxes in the range $6.3-7.0$ keV.

There are fundamental differences between what we present here and previous disk reprocessing codes. First, we use plane-parallel slabs of \emph{finite} thickness which are illuminated by the corona on both sides, whereas it is usually assumed that the disk is \emph{semi-infinite} [cf. {\sc reflion} \citep{ros05a}, the {\sc xillver} code of \citet{gar10a, gar11a, gar13a}, and its extension {\sc relxill} \citep{gar14a}]. This is an important distinction, because the total optical depth of the disk drops to $\lesssim 1$ in its inner region, where the radiative diffusion approximation, which typically supplies the lower boundary condition for transfer calculations [e.g., a blackbody, with the radial temperature dependence of \citet{sha73a}, placed at some large optical depth beneath the disk surface], is not valid. In contrast, our upper and lower boundary conditions---the flux incident upon the upper and lower surfaces of the disk---remain physical even when the disk becomes optically thin. Second, both the density structure of our plane-parallel slabs and the flux irradiating their surfaces are computed using realistic dynamic models instead of assumed on the basis of simple and often arbitrary analytic relations. Since inhomogeneities in the density structure can have important effects on Fe K$\alpha$ production \citep{bal04a}, drawing it from an MHD simulation represents a particularly significant improvement.

Using the ray-tracing code {\sc pandurata}, the Fe K$\alpha$ photons are transported from the disk surface to an observer at infinity. This geodesic transport includes all special and general relativistic effects, as well as returning radiation [photons that are deflected by the black hole's gravity and then scatter off the disk; see \citep{sch09a}] {  and Compton scattering off coronal electrons.}

For future calculations with AGN parameters, we expect X-ray absorption in the disk to become an important effect. We plan to implement an iterative scheme between {\sc pandurata} and {\sc ptransx}, in which the absorbed hard X-ray flux will be reprocessed into additional seed photons available for upscattering in the corona. In this way we will achieve a truly global (disk and corona) self-consistent radiation field. Toward this end, we also plan to replace the \emph{ad hoc} temperature floor component of our calculation with a more physical approach: we will interpret the local cooling rate from the MHD simulation as a local heating rate due to dissipation of MHD turbulence in order to calculate the local temperature balance in the disk body.

\section{Results}

\subsection{Fe K$\alpha$ Emission in the Fluid Rest Frame}

\begin{figure}[H]
\begin{center}
\includegraphics[width=0.8\textwidth]{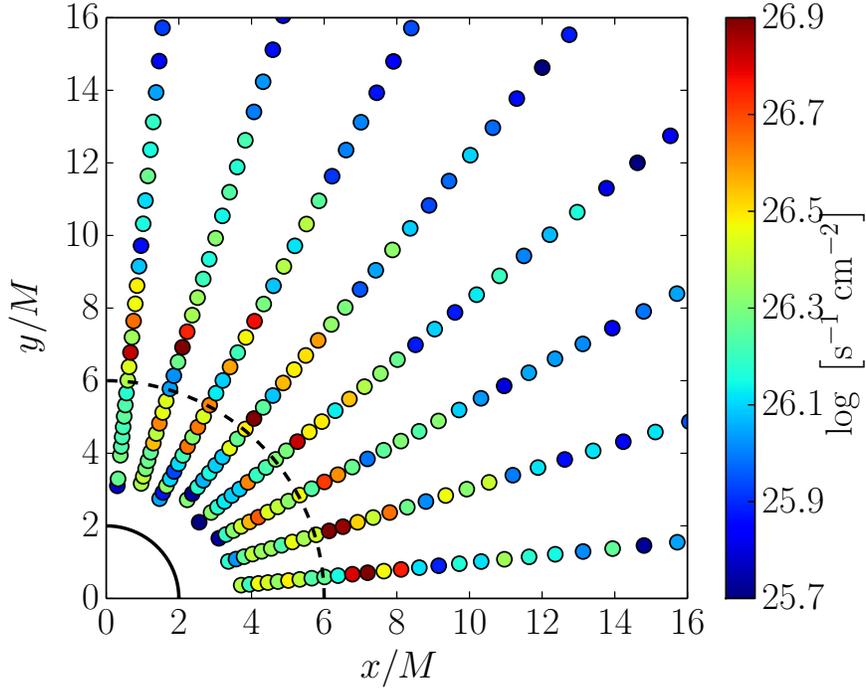}
\caption{A sampling of the Fe K$\alpha$ surface brightness in the fluid rest frame on the upper disk surface for one quadrant of the accretion disk about a $10 M_\odot$ Schwarzschild black hole, with an accretion rate of $1\%$ the Eddington value. Not all $(r, \phi)$ points are shown in this view. The inner solid black line represents the location of the event horizon and the outer dashed black line represents the radius of the ISCO. The reverse view---looking up toward the midplane---is, on this scale, nearly indistinguishable.}
\label{feka_surf}
\end{center}
\end{figure}

\begin{figure}[H]
\begin{center}
\includegraphics[width=0.8\textwidth]{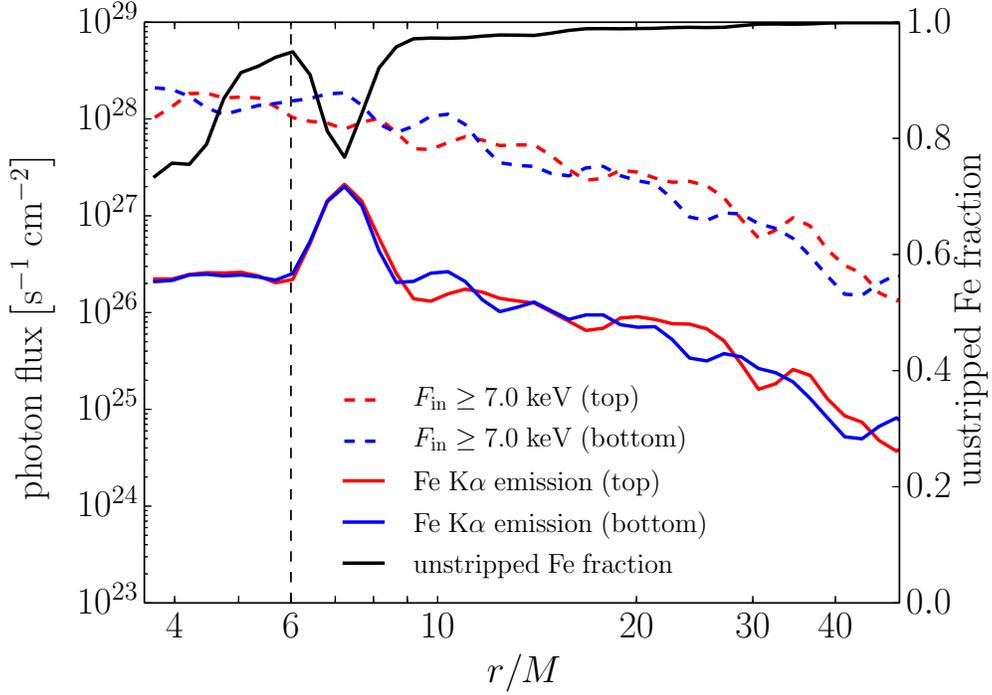}
\caption{The red and blue solid lines are the Fe K$\alpha$ surface brightness in the fluid rest frame for the $\phi = 0$ azimuth as a function of radius, for the upper and lower disk surfaces, respectively. The dashed red and blue lines are the photon flux incident upon the upper and lower disk surfaces, respectively, integrated above 7.0 keV. These four lines correspond to the left axis. The black line is the fraction of unstripped Fe atoms in the disk at $\phi = 0$ as a function of radius, corresponding to the right axis. The vertical dashed line indicates the ISCO. To facilitate interpretation, all curves have had a Gaussian smoothing kernel applied, with standard deviation equal to the radial cell spacing.}
\label{feka_rad_profile}
\end{center}
\end{figure}

\begin{figure}[H]
\begin{center}
\includegraphics[width=0.8\textwidth]{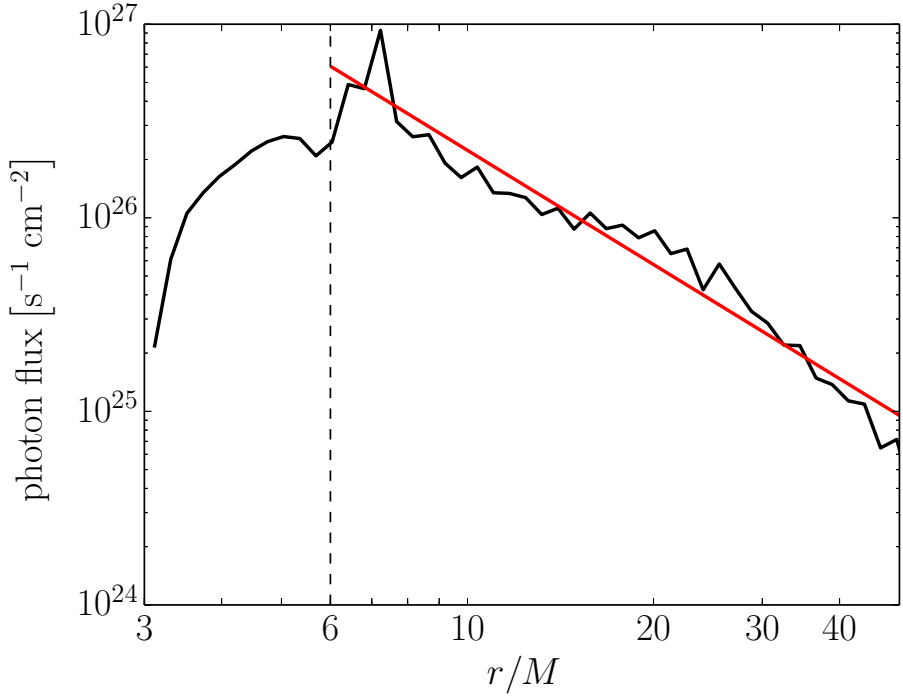}
\caption{The black line is the Fe K$\alpha$ surface brightness, averaged over the top and bottom of the disk and over azimuth. No smoothing kernel has been applied. The red line is a power-law fit to the region beyond $r \simeq 7M$. The fitted power-law index is $-2$. As in Figure \ref{feka_rad_profile}, the dashed line indicates the radius of the ISCO.}
\label{power_law}
\end{center}
\end{figure}

Our Fe K$\alpha$ surface brightness predictions are summarized in three figures. Figure \ref{feka_surf} shows the Fe K$\alpha$ surface brightness in the fluid rest frame for one quadrant of the accretion disk at a sampling of 394 $(r, \phi)$ points [8 evenly spaced azimuthal zones, each with $48 - 51$ logarithmically ($\Delta r/r = 0.06$) spaced radial zones; the number of radial zones per azimuth varies depending on density variations at small radii (see Figure \ref{feka_surf})] for one snapshot in time of our $10 M_\odot$, Schwarzschild, $1\%$ Eddington case, assuming solar Fe abundance. The solid red line in Figure \ref{feka_rad_profile} presents the Fe K$\alpha$ surface brightness for the $\phi = 0$ azimuthal slice of the same data (after the application of a smoothing kernel). The most salient feature of Figure \ref{feka_rad_profile} is the rough power-law in radius the Fe K$\alpha$ emission appears to obey exterior to a maximum occurring at $\simeq 7M$, $\simeq 1 M$ outside the ISCO. This behavior is more apparent in the $\phi$-averaged picture of Figure \ref{power_law}, where we note that the the power-law portion of the $\phi$-averaged Fe K$\alpha$ surface brightness varies with radius with index $-2$. The steeper than $\propto r^{-2}$ decline from $7$ to $8M$ seen in both Figure \ref{feka_rad_profile} and \ref{power_law} is due to two effects. First, as the disk becomes optically thick, some K$\alpha$ photons are absorbed in the disk body before they escape. This also explains why the Fe K$\alpha$ surface brightness of the upper and lower surfaces (the red and blue curves of Figure \ref{feka_rad_profile}) vary together before $\simeq 8M$, but are independent at larger radii. In addition, the atomic fluorescence yield of Fe is dependent upon ionization state (see below), and the higher ionization found near $7M$ has an effectively higher atomic fluorescence yield than the less-ionized Fe at larger radii.

To understand why the decline in Fe K$\alpha$ surface brightness at small radii occurs, consider the black line in Figure \ref{feka_rad_profile}, representing the fraction of Fe atoms in the disk at that radius and azimuth retaining at least one electron. At smaller radii, the disk's surface density decreases, and the ionization parameter increases; within about $10 M$, the fraction of unstripped Fe begins to decrease. Though recombination onto bare Fe nuclei still provides a fluorescence mechanism in the highly ionized gas, the recombination rate is proportional to the unstripped Fe fraction, and so Fe K$\alpha$ emission through recombination decreases at small radii as well. {  In addition, as the disk thins, its column density decreases---there are simply fewer Fe atoms to undergo fluorescence at small radii.}

\begin{figure}[H]
\begin{center}
\includegraphics[width=0.8\textwidth]{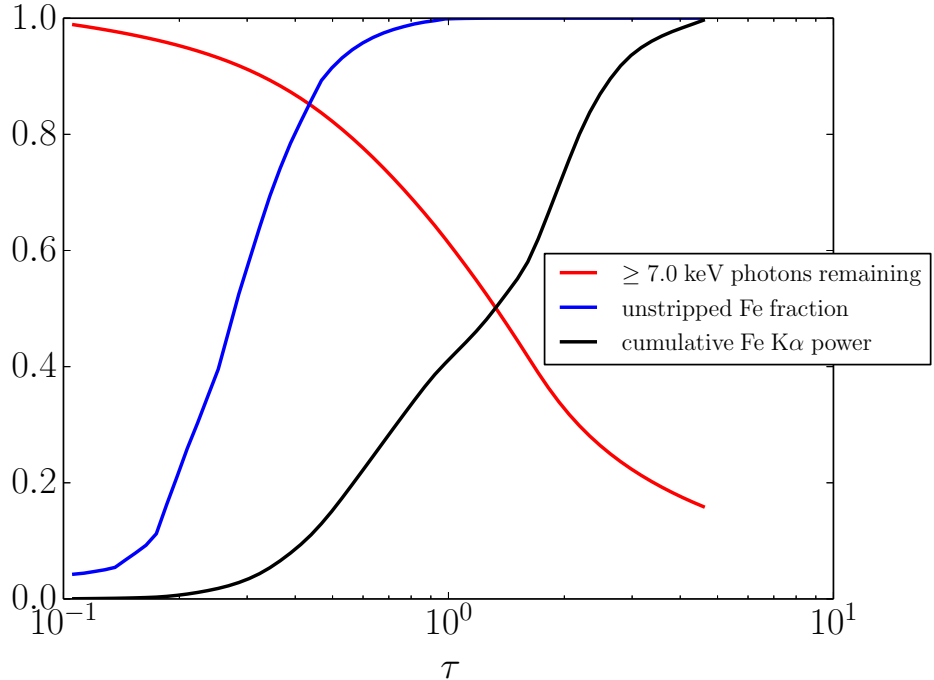}
\caption{Cumulative Fe K$\alpha$ (relative to the total Fe K$\alpha$ power in the lower half of the disk), unstripped Fe fraction, and fraction of incident flux (above 7.0 keV) remaining as a function of optical depth at $\phi = 0$, $r = 10M$. The lines terminate at the midplane for this azimuth and radius.}
\label{yield}
\end{center}
\end{figure}

Figure \ref{feka_rad_profile} also shows the photon flux incident upon the disk surface, integrated above 7.0 keV. The K-edge---the photon energy required to induce K$\alpha$ fluorescence---varies with Fe ionization state, from approximately 7.1 keV for neutral Fe to 9.3 keV for H-like Fe \citep{kal04a}. The fluorescence yield---the fraction of absorbed K-edge photons resulting in the production of a K$\alpha$ photon---varies with ionization state as well \citep{kro87a}, but is typically $\sim 0.5$. It is at first surprising, then, that while the Fe K$\alpha$ surface brightness in Figure \ref{feka_rad_profile} roughly follows the flux of incident photons above the K-edge---at least exterior to the peak at about $7 M$---the former is approximately two orders of magnitude smaller than the latter, smaller than the factor of 0.5 we might expect from the fluorescence yield. Figure \ref{yield} provides a physical explanation. Due to the disk's high degree of ionization, Thomson scattering occurs throughout its volume, but appreciable Fe K$\alpha$ production can occur only in those regions in which there is a sufficient population of unstripped Fe atoms. However, on much of the disk's surface, the incident flux is so large that a significant number of unstripped Fe atoms can exist only at optical depths where the flux has been substantially reduced by reflection. For the particular $(r, \phi)$ slab considered in Figure \ref{yield}, we see that after one optical depth, at which the fraction of unstripped Fe reaches unity, approximately $40\%$ of the photons incident upon the disk with energy above the K-edge have been absorbed or scattered out. This first optical depth accounts for slightly less than half of the total Fe K$\alpha$ photons produced in (the lower half of) this slab, but for a greater portion of the K$\alpha$ \emph{emission}, since K$\alpha$ photons produced at points deeper must diffuse through an optically thick disk body before escaping. Thus, the reflective outer layers of the accretion disk act to suppress K$\alpha$ production, depressing the \emph{effective} fluorescent yield relative to the \emph{atomic} fluorescent yield. \citet{nay00a} and \citet{gar10a} note a similar ``hot skin'' effect, where the first optical depth or so of their irradiated slabs are at a much higher temperature and ionization than the underlying material. For the high ionization parameter cases of \citet{gar10a}, they report a marked decrease in the reprocessed Fe K$\alpha$ equivalent width, generally consistent with our ``depressed effective yield'' interpretation.

Figure \ref{yield} also provides the justification for our choice to forego a redistribution function which accurately describes Compton scattering of Fe K$\alpha$ lines in lieu of our simpler scheme. From Figure \ref{yield}---which is qualitatively similar for any $(r, \phi)$ point exterior to $\simeq 7M$---we see that the median Fe K$\alpha$ photon is produced at an optical depth $\sim 1$ below the disk surface, and therefore experiences on average one Compton scattering event as it escapes. Typically, $kT \sim 5$ keV in this region, so for Fe K$\alpha$ photons with energy $6.3 - 7.0$ keV, the broadening due to Compton scattering, $\sim 2\%$, is small. {  It should be noted that the effects of Compton heating and cooling are included in {\sc xstar}'s determination of the temperature, though the radiation field computed by {\sc ptransx} does not reflect this; future versions of this code---which we will use to compute the reprocessed continuum in addition to line emission---will include a Compton scattering kernel.}

\subsection{Fe K$\alpha$ Line Profiles}

\begin{figure}[H]
\begin{center}
\includegraphics[width=0.8\textwidth]{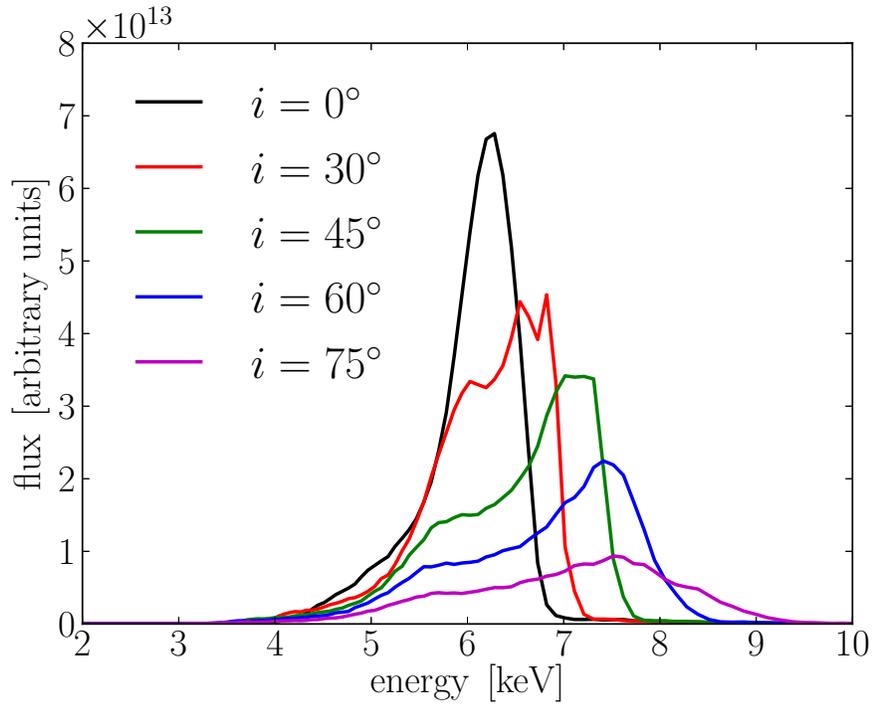}
\caption{The Fe K$\alpha$ line profile as would be seen by a distant observer, at several inclinations $i$ (where $0^\circ$ is viewing the disk face-on), excluding electron scattering of the line photons as they travel through the corona.}
\label{feka_line_profile}
\end{center}
\end{figure}

\begin{figure}[H]
\begin{center}
\includegraphics[width=0.8\textwidth]{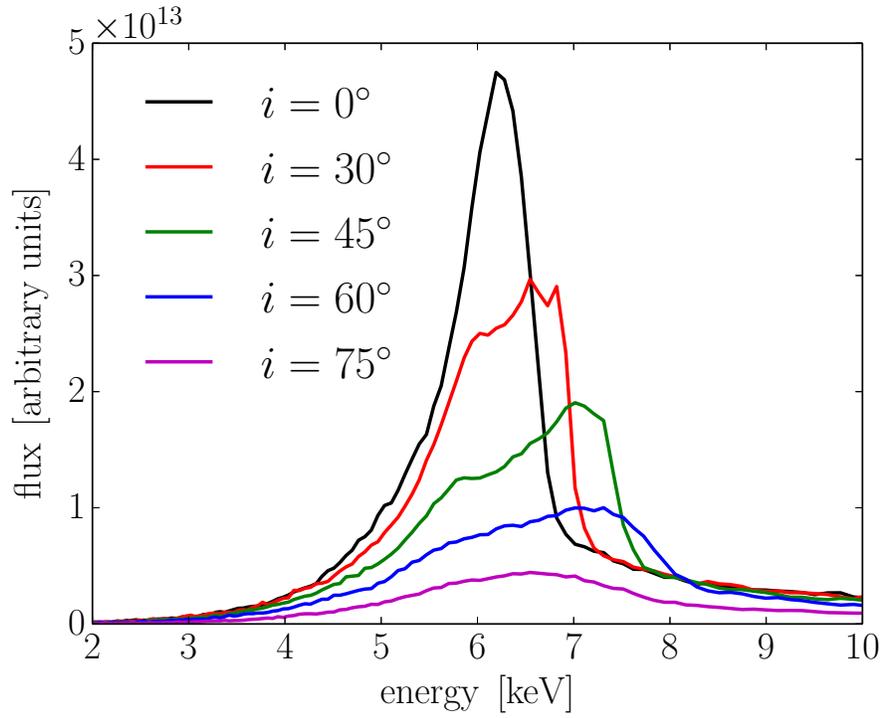}
\caption{The Fe K$\alpha$ line profile as would be seen by a distant observer at several inclinations, but including the effects of electron scattering of the line photons in the hot corona.}
\label{feka_line_profile_compton}
\end{center}
\end{figure}

Figure \ref{feka_line_profile} shows the shape of the Fe K$\alpha$ lines as measured by a distant observer at several inclinations, including all special and general relativistic effects, {  but without electron scattering of line photons in the corona; Figure \ref{feka_line_profile_compton} shows the line profiles with this effect included}. The line profiles in both Figures \ref{feka_line_profile} and \ref{feka_line_profile_compton} possess features similar to those from actual observations: the emission line is strongly and asymmetrically broadened, and at non-zero inclinations takes on a ``double-horned'' quality [compare, e.g., to the observational results discussed in \citet{mil07a}]. {  As might be expected, Compton scattering in the hot corona tends to redistribute line photons to higher energies.  Such scatters are relatively few for photons initially directed upward from the disk, but they are much more numerous for photons that might otherwise go to observers closer to the disk plane.  The result is to alter only slightly the line profiles seen by observers viewing the disk face-on, but to diminish the equivalent width of and broaden the line that observers with a more edge-on view might see [see \citet{wil15a}].}

\begin{figure}[H]
\begin{center}
\includegraphics[width=0.8\textwidth]{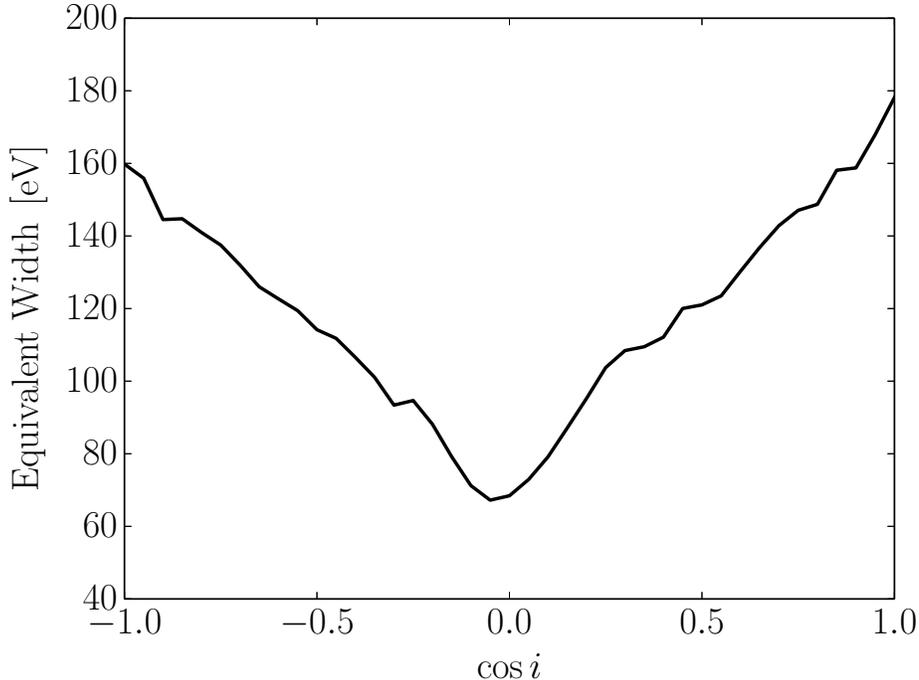}
\caption{The equivalent width of the Fe K$\alpha$ line as a function of observer inclination angle, including the effects of Compton scattering of the line photons through the corona.}
\label{ew}
\end{center}
\end{figure}

The equivalent width of the Fe K$\alpha$ line (as would be measured by a distant observer, {  and including coronal Compton scattering}) as a function of inclination angle is presented in Figure \ref{ew}. The equivalent widths we find, in the range $60-180$ eV, are in agreement both with typical observational values [see \citet{rey03a}] and previous disk reprocessing codes (namely {\sc reflion}, {\sc xillver}, and {\sc relxill}, as discussed above). It is important to note that the specific equivalent widths we report are for solar Fe abundance, though we expect the equivalent width to vary roughly linearly with Fe abundance. The dip centered at $i = 90^\circ$---viewing the disk edge-on---is due to obscuration of the disk surface by the disk itself {  as well as losses due to Compton scattering in the corona}. Due to gravitational lensing, even edge-on observers receive some of the line emission from the disk, and so the equivalent width does not go to zero at $i = 90^\circ$.

\section{Discussion}

Figures \ref{feka_line_profile}, \ref{feka_line_profile_compton}, and \ref{ew} represent the culmination of our efforts so far. As we have stressed throughout the development above, our prediction of observed Fe K$\alpha$ lines requires as input only a small number of physical parameters: the black hole mass and spin, the accretion rate, the iron abundance, and the observer inclination. Significantly, none of these parameters stands in for an unknown physical process. The most important aspect of Figure \ref{feka_line_profile} is that it represents a quantitative \emph{prediction} for how we expect the Fe K$\alpha$ line profile to appear from a $10 M_\odot$  Schwarzschild black hole accreting at $1\%$ the Eddington value, predicated on fundamental physics. We combine global MHD simulations performed in full GR, a Monte Carlo radiation transfer code to determine the electron temperature throughout the corona and the hard X-ray flux irradiating each point on the disk surface, and the disk reprocessing code we describe above to construct maps of the Fe K$\alpha$ emission over the disk surface. The final results---Fe K$\alpha$ line profiles---are observables founded upon the physics which describe accreting black hole systems. The most significant gap remaining in the physics is the equation of state used by the MHD code. The state-of-the-art for simulations of bright, but sub-Eddington, accretion onto black holes assumes a local optically thin cooling function designed to radiate quickly nearly all the heat dissipated, maintaining a disk scale height as a function of radius determined {\it a priori}. {  For the simulation used here, this scale height profile is close to what might be expected from conventional analytic accretion disk theory \citep{sha73a} assuming a radiation-dominated disk with $\dot{m} \simeq 0.2$; in forthcoming work, we will explore how the Fe K$\alpha$ properties presented here depend on adjusting our choice of $\dot{m}$. \citet{sch13a} present a more detailed account of the relation between analytic accretion disk theory and the terms of our simulation.} As simulation codes coupling radiation transfer to MHD \citep{jiang14, sadow14} become more efficient and eliminate restrictive approximations, it will be possible to close this gap.

Model fits to data typically assume that the Fe K$\alpha$ surface brightness is axisymmetric and follows a power-law in radius with a cutoff at the ISCO. From Figure \ref{power_law}, it is clear that, for this particular example, our more careful approach validates this model to some extent: at least for the one point in the black hole parameter space we have considered, the radial variation in Fe K$\alpha$ emission turns out to be roughly power-law. The fundamental difference is that the power-law index we find, $-2$, the location of the maximum, $r \simeq 7M$, and the very fact that the behavior is power-law at all, arise in a natural way from the calculation: they are neither fits to observations nor put in ``by hand.''

The power-law we find is, in fact, shallower than those typically used in phenomenological models, but this is a direct consequence of our \emph{extended} coronal geometry. For radii greater than that of the peak surface brightness, the Fe K$\alpha$ emission should vary like the incident hard X-ray flux---as in Figure \ref{feka_rad_profile}. It is easily shown (in flat space) that if the corona is treated as a point source at some height $z$ above the center of a disk of constant aspect ratio, the variation of the flux with the cylindrical radial coordinate $r$ is either proportional to $r^{-3}$ for the case when $z \ll r$, or independent of $r$ in the case when $z \gg r$. Fully relativistic ray-tracing calculations of ``lamppost" geometries, like those of \citet{wil12a} and \citet{dau13a}, typically find steep power-laws in the inner disk regions, where $z \gg r$, leveling off to $r^{-3}$ at large radii, where $z \ll r$. When they investigate more extended, but still arbitrarily chosen, hard X-ray emissivity distributions \citep{wil12a}, the K$\alpha$ emissivity in the region covered by the extended hard X-ray source roughly mirrors the coronal emissivity. Our {\it physically-derived} result that the K$\alpha$ emissivity is $\propto r^{-2}$ is therefore a direct consequence of the similarly extended coronal emission that follows directly from the underlying MHD simulation. It is, however, important to note that the emissivity profile may change as a function of accretion rate, black hole spin, or black hole mass. Magnetic field topology may also be important, as the driver behind different coronal and jet properties \citep{bec08a}.

\begin{figure}[H]
\begin{center}
\includegraphics[width=0.8\textwidth]{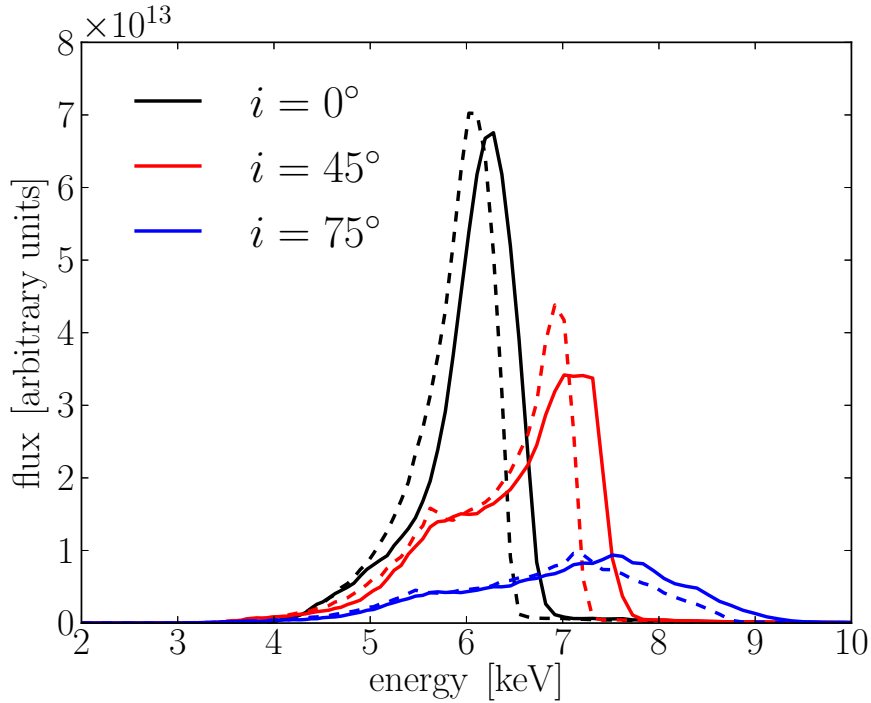}
\caption{The Fe K$\alpha$ line profile as would be seen by a distant observer at several inclinations: those in dashed lines were computed from the power-law fit extended back to the ISCO (assuming 6.4 keV Fe K$\alpha$ photons), e.g., the red line in Figure \ref{power_law}, while those in solid lines are reproduced from Figure \ref{feka_line_profile} for comparison; for both solid and dashed lines, color indicates inclination.}
\label{feka_line_profile_plfit}
\end{center}
\end{figure}

Of particular importance to the use of observed Fe K$\alpha$ line profiles to infer black hole spin is the fact that we find the line emission peaks approximately $1 M$ \emph{outside} the ISCO. We have, of course, not performed a sufficiently detailed exploration of the black hole parameter space to determine whether this is a systematic effect---indeed, it is possible that for a different set of parameters, the peak may be found \emph{inside} the ISCO. This latter alternative might, for example, be expected when the accretion rate (and thus disk surface density) is higher \citep{sch13a}. In addition, in generating the line profiles of Figures \ref{feka_line_profile} and \ref{feka_line_profile_compton}, we keep track of three separate energy channels within the $6.3-7.0$ keV range; higher ionization states of Fe produce K$\alpha$ photons at greater energies, and this potentially several-hundred eV difference can have significant effects on the predicted line profile. As an illustration, we show in Figure \ref{feka_line_profile_plfit} the Fe K$\alpha$ line profiles seen by a distant observer using the power-law fit extended back to the ISCO from Figure \ref{power_law}, assuming all Fe K$\alpha$ emission occurs at 6.4 keV (as is often done), compared to those generated by our method, i.e., from Figure \ref{feka_line_profile}. By moving the location of the interior cutoff inwards by $M$, removing physically important features---like the brief yet steep decline in flux from $7$ to $8M$ discussed in the previous section---which are \emph{not} captured by a pure power-law, and assuming all K$\alpha$ photons are produced by near neutral Fe, the line profile is reddened and altered in shape. {  When comparing to observed spectra, this shift can be expected to have particularly important consequences for constraining the system's inclination, for example.}

By further exploration of the parameter space, we will learn how variations in mass, spin, accretion rate, and Fe abundance manifest themselves in the strength and shape of the Fe K$\alpha$ line profile. Ultimately, we envision the construction of a grid of model profiles spanning the parameter space, forming the foundation for an extension to {\sc xspec} \citep{arn96a} which will take as input an X-ray spectrum of a stellar-mass black hole or AGN from an observatory such as \emph{Chandra}, \emph{XMM-Newton}, \emph{Suzaku}, or \emph{NuSTAR}, and will output the best-fit values of the intrinsic parameters, derived from physically-based, rather than phenomenological, models.

\acknowledgments

This work was partially supported by NASA/ATP Grant NNX14AB43G, NSF Grant AST-0908869, and NASA/ATP Grant 13-0077. We are particularly grateful to John Hawley for providing funds from the NSF grant and to John Baker for funds from the latter NASA grant. BEK also thanks the GSFC Laboratory for High Energy Astrophysics for hospitality.

\bibliography{references}

\end{document}